\documentstyle[12pt]{article}
\newcommand{\inv}{\left. g^0_A \right|_{\rm inv}}

\newcommand{\Frac}[2]%
{{\textstyle \frac{\mbox{\footnotesize $#1$}\rule[-0.9mm]{0mm}{1mm}}%
{\mbox{\footnotesize $#2$}\rule{0mm}{3.1mm}}}}
\renewcommand{\thefootnote}{\fnsymbol{footnote}}

\begin{document}
\begin{titlepage}
\vspace*{-12 mm}
\noindent
\begin{flushright}
\begin{tabular}{l@{}}
TUM/T39-99-15 \\
hep-ph/9907373 \\
\end{tabular}
\end{flushright}
\vskip 12 mm
\begin{center}
{\Large \bf Gluons and the $\eta'$--nucleon coupling constant
\footnote[1]{Work supported in part by BMBF.} } 
\\[10mm]
{\bf Steven D. Bass}
%%\footnote{sbass@physik.tu-muenchen.de} 
\\[10mm]   
{\em 
Physik Department, Technische Universit\"at M\"unchen, \\
D-85747 Garching, Germany }

\end{center}
\vskip 10 mm
\begin{abstract}
\noindent
We derive the effective chiral Lagrangian for low-energy $\eta$--nucleon 
and $\eta'$--nucleon interactions and show that gluonic degrees of 
freedom, via the axial $U_A(1)$ anomaly, induce a contact term in the
$pp \rightarrow pp \eta$ and $pp \rightarrow pp \eta'$ reactions.
We then discuss the consequences for the extraction of $g_{\eta' NN}$ 
from experimental data.
\end{abstract}
\end{titlepage}
\renewcommand{\labelenumi}{(\alph{enumi})}
\renewcommand{\labelenumii}{(\roman{enumii})}
\renewcommand{\thefootnote}{\arabic{footnote}}
\newpage
%
%\baselineskip=6truemm

\section{Introduction}

Polarised deep inelastic scattering \cite{bass99,windmolders} and $\eta'$ 
physics \cite{christos} provide complementary windows on dynamics induced 
by the axial $U_A(1)$ anomaly \cite{zuoz} in QCD. 
The flavour-singlet Goldberger-Treiman relation \cite{venez} relates the 
flavour-singlet axial-charge $g_A^{(0)}$ measured in polarised deep inelastic 
scattering to the $\eta'$--nucleon coupling constant $g_{\eta' NN}$.
The large mass of the $\eta'$ and the small value of $g_A^{(0)}$ extracted
from deep inelastic scattering point to substantial violations of the OZI 
rule \cite{ozi} in the flavour-singlet $J^P=1^+$ channel \cite{okubo}.
In this paper we discuss the role of gluonic degrees of freedom in the
low-energy $\eta'$--nucleon interaction. We derive the effective chiral 
Lagrangian for $\eta'$--nucleon interactions and explain why gluons 
induce a contact term in the low-energy $pp \rightarrow pp \eta'$ 
reaction.
The strength of this contact term is, in part, related to the amount of 
spin carried by polarised gluons in a polarised proton. The total cross 
section for $pp \rightarrow pp \eta'$ near threshold has been measured at 
COSY \cite{cosy} and SATURNE \cite{saturne}. New measurements could follow 
from an upgraded CELSIUS machine. The $\eta'$ photoproduction process is 
being studied in experiments at ELSA \cite{elsa} and Jefferson Laboratory 
\cite{cebaf}.

We begin in Section 2 with a brief review of the effective Lagrangian 
\cite{witten,vecca,lagran} for low-energy $\eta'$--meson interactions.
In Section 3 we extend this theory to include $\eta'$--nucleon coupling. 
Finally, in Section 4, we discuss the possible size of gluonic effects
in the $\eta'$--nucleon interaction and the consequences for the extraction 
of $g_{\eta' NN}$ from experimental data.

\section{The low-energy effective Lagrangian}

Starting in the meson sector, the effective Lagrangian 
\cite{witten,vecca,lagran} for low-energy QCD is constructed as follows.
We begin with the SU(3) chiral Lagrangian 
\begin{equation}
{\cal L} = 
{F_{\pi}^2 \over 4} {\rm Tr}(\partial^{\mu}U \partial_{\mu}U^{\dagger}) 
+
{F_{\pi}^2 \over 4} 
{\rm Tr} M \ \biggl( U + U^{\dagger} \biggr)
\end{equation}
where $U$ is the unitary meson matrix and
$M = 
{\rm diag} [ m_{\pi}^2, m_{\pi}^2, (2 m_K^2 - m_{\pi}^2 ) ]$
is the meson mass matrix
--- for a review see \cite{houches}. 
The matrix $U$ is extended to include a flavour-singlet Goldstone
boson \cite{weinberg}:
\begin{equation}
U = \exp \ \biggl(  i {\phi \over F_{\pi}}  
                  + i \sqrt{2 \over 3} {\eta_0 \over F_0} \biggr) .
\end{equation}
Here 
\begin{equation}
\phi = \ \sqrt{2}
\left(\begin{array}{ccc} 
{1 \over \sqrt{2}} \pi^0 + {1 \over \sqrt{6}} \eta_8 & \pi^+ & K^+ \\
\\
\pi^- & 
-{1 \over \sqrt{2}} \pi^0 + {1\over \sqrt{6}} \eta_8 & K^0 \\
\\
K^- & {\bar K}^0 & -{2 \over \sqrt{6}} \eta_8 
\vphantom{\inv}  
\end{array}\right) 
\end{equation}
denotes the octet of would-be Goldstone bosons associated with 
spontaneous chiral $SU(3)_L \otimes SU(3)_R$ breaking and $\eta_0$ 
is the singlet boson.
The pion decay constant $F_{\pi} = 92.4$MeV;
$F_0$ renormalises the flavour-singlet decay constant --- see Eq.(8) below.

Next, one introduces $U_A(1)$ terms to generate a gluonic mass term for 
the $\eta_0$ and to reproduce the anomaly \cite{adler,bell} in the 
divergence of the gauge-invariantly renormalised flavour-singlet 
axial-vector current
\begin{equation}
J_{\mu 5} =
\left[ \bar{u}\gamma_\mu\gamma_5u
                  + \bar{d}\gamma_\mu\gamma_5d
                  + \bar{s}\gamma_\mu\gamma_5s \right]_{GI}^{\mu^2} ,
\end{equation} 
viz.
\begin{equation}
\partial^\mu J_{\mu5} = 
\sum_{k=1}^{f} 2 i \biggl[ m_k \bar{q}_k \gamma_5 q_k 
\biggr]_{GI}
+ N_f 
\biggl[ {\alpha_s \over 4 \pi} G_{\mu \nu} {\tilde G}^{\mu \nu} 
\biggr]_{GI}^{\mu^2}.
\end{equation}
Here $N_f=3$ is the number of light flavours, the subscript
$GI$ denotes
gauge invariant renormalisation and the superscript $\mu^2$ 
denotes the renormalisation scale.
The Adler-Bardeen theorem \cite{bardeen} states that the anomaly on 
the right hand side of Eq.(5) is not renormalised to all orders in 
perturbation theory. We use this theorem to constrain the possible 
$U_A(1)$ breaking terms in the effective Lagrangian.

The low-energy effective Lagrangian \cite{vecca,lagran} is
\begin{eqnarray}
{\cal L}_{\rm m} = & &
{F_{\pi}^2 \over 4} 
{\rm Tr}(\partial^{\mu}U \partial_{\mu}U^{\dagger}) 
+
{F_{\pi}^2 \over 4} {\rm Tr} M \ \biggl( U + U^{\dagger} \biggr)
\\ \nonumber
&+& 
 b {F_{\pi}^2 \over 4} 
 \biggl( {\rm Tr} U^{\dagger} \partial_{\mu} U \biggr)^2
+ {1 \over 2} i Q {\rm Tr} \biggl[ \log U - \log U^{\dagger} \biggr]
+ {1 \over a F_{0}^2} Q^2 .
\end{eqnarray}
The gluonic term 
$Q = {\alpha_s \over 4 \pi} G_{\mu \nu} {\tilde G}^{\mu \nu}$ 
is treated as a background field. 
It has no kinetic term
\footnote{In QCD $Q = \partial^{\mu} K_{\mu}$ where $K_{\mu}$ is the 
          gauge-dependent gluonic Chern-Simons current.
          The reader may wish to think of $Q^2$ as a kinetic term for 
          $K_{\mu}$.}
and mixes with the $\eta_0$ to generate a gluonic mass term for the 
$\eta'$ --- see below.
The term proportional to $b$ is fourth order in $U$ but gives a second 
order term in $\eta_0$.
For simplicity we omit fourth order terms in the meson fields.
Their inclusion is straightforward but of no primary influence
in our context.
The most general low-energy effective Lagrangian involves a $U_A(1)$
invariant polynomial in $Q^2$.  Higher-order terms in $Q^2$ become 
important when we consider scattering processes involving more than
one $\eta'$ \cite{veccb}.

The $U_A(1)$ transformation for the effective Lagrangian (6) is defined 
\cite{christos} by 
$U \rightarrow \exp {(2 i \beta)} \ U$; 
$Q$ is treated as $U_A(1)$ invariant. 
The flavour-singlet axial-vector current 
\begin{equation}
J_{\mu 5} = \sqrt{6} F_{\rm singlet} \partial_{\mu} \eta_0
\end{equation}
with
\begin{equation}
F_{\rm singlet} = 
{ F_{\pi}^2 \over F_0 } \biggl( 1 - 3 b \biggr)
\end{equation}
satisfies the anomalous divergence equation 
$\partial^{\mu} J_{\mu 5} = N_f Q + \ {\rm mass \  terms}$.
The flavour-singlet decay constant is renormalised 
relative to $F_{\pi}$ by gluonic intermediate states
($q {\bar q} \rightarrow gg \rightarrow q {\bar q}$).
The parameters $F_0$ and $b$ describe two distinct 
renormalisation effects:
$F_0$
contributes to $F_{\rm singlet}$ and also the meson masses  
(through its appearance in ${\rm Tr} M (U + U^{\dagger})$)
whereas $b$ contributes only to $F_{\rm singlet}$.
Assuming a continuous large $N_c$ limit, 
\begin{equation}
\lim_{N_c \rightarrow \infty} F_{\rm singlet} = F_{\pi}
\end{equation}
in the chiral limit.

The gluonic term $Q$ can be eliminated from the effective Lagrangian
through its equation of motion.
Expanding to ${\cal O}(p^2)$ in momentum and keeping finite quark
masses one finds
\begin{eqnarray}
{\cal L}_{\rm m} = & &
{1 \over 2} \partial^{\mu} \pi_a \partial_{\mu} \pi_a 
+
{1 \over 2} \partial_{\mu} \eta_0 \partial^{\mu} \eta_0 
\ \biggl( {F_{\pi} \over F_0} \biggr)^2 \ ( 1 - 3 b )
\\ \nonumber
&-& 
{3 a \over 2} \eta_0^2 \ 
\\ \nonumber
&-& {1 \over 2} 
m_{\pi}^2 \biggl( 2 \pi^+ \pi^- + \pi_0^2 \biggr)
- m_K^2 \biggl( K^+ K^- + K^0 {\bar K}^0 \biggr)
- {1 \over 2} \biggl( {4 \over 3} m_K^2 - {1\over 3} m_{\pi}^2 \biggr)
  \eta_8^2
\\ \nonumber
&-& {1 \over 2} \biggl( {2 \over 3} m_K^2 + {1\over 3} m_{\pi}^2 \biggr)
    \ \biggl( {F_{\pi} \over F_0} \biggr)^2 \ \eta_0^2
    + {4 \over 3 \sqrt{2}} \biggl( m_K^2 - m_{\pi}^2 \biggr) 
    \ \biggl( {F_{\pi} \over F_0} \biggr) \eta_8 \eta_0
\end{eqnarray}
where $\phi_a$ with $a=1,...,8$ refers to the octet Goldstone boson fields.
In the chiral limit $M=0$ gluons contribute a finite mass 
\begin{equation}
m_{\eta_0}^2 = {3a \over \biggl( {F_{\pi} \over F_0} \biggr)^2 \ ( 1 - 3 b )}
\end{equation}
to the singlet $\eta_0$.
The masses of the physical $\eta$ and $\eta'$ mesons 
are found by diagonalising the $(\eta_8, \eta_0)$ mass matrix
which follows from Eq.(10).

Henceforth, we take 
$F_0 \sim 0.1$GeV \cite{gilman} and 
${\tilde m}_{\eta_0}^2 \equiv 3 a \sim 1$GeV$^2$.

It is worthwhile to comment on the behaviour of ${\cal L}_{\rm m}$
when we take $N_c$, the number of colours, to infinity.
In QCD the axial anomaly decouples as $1/N_c$ when we take 
$N_c \rightarrow \infty$. 
This is reflected in the equation for the $\eta_0$ mass squared if 
we take 
$a \propto 1/N_c$ \cite{witten}.
Phenomenologically, the large mass of the $\eta'$ 
($m_{\eta'}  \sim 1$GeV) 
means that OZI and large $N_c$ are not always good approximations 
in the $U_A(1)$ channel.
A second source of OZI violation is the non-vanishing 
anomalous dimension \cite{koeb,rjc,kod} of the flavour-singlet 
axial vector current (4). 
In QCD $J_{\mu 5}$ satisfies the renormalisation group equation
\begin{equation}
J_{\mu 5} (\lambda) = J_{\mu 5} (\infty) / E [ \alpha_s(\lambda) ]
\end{equation}
where
\begin{equation}
E [ \alpha_s (\lambda) ] 
= \exp \int^{\alpha_s(\lambda)}_0 
\! d{\tilde \alpha_s}\, 
\gamma({\tilde \alpha_s})/\beta({\tilde \alpha_s}) .
\end{equation}
Here
$\gamma(\alpha_s)$
($= f {\alpha_s^2 \over \pi^2} + {\cal O}(\alpha_s^3)$) 
is the (two loop) non-zero anomalous dimension of 
$J_{\mu 5}$ and $\beta (\alpha_s)$ is the QCD beta function.
We are free to choose the QCD coupling $\alpha_s(\lambda)$ at either 
a hard or a soft scale $\lambda$.
If we work to $O(\alpha_s^2)$ in perturbation theory, 
then 
$E [ \alpha_s ] \sim 0.84$ when  $\alpha_s \sim 0.6$, 
typical of the infrared
-- see eg. \cite{bass99}.
At least in perturbation theory, $E [ \alpha_s ]$ remains close 
to the OZI value $E [ \alpha_s ] = 1$ ---
in contrast to $m_{\eta_0}^2$ which exhibits large OZI violation. 
For the meson theory (6) the renormalisation group factor $E[ \alpha_s ]$ 
can be absorbed in the parameter $b$.

\section{The $\eta'$--nucleon interaction}

The low-energy effective Lagrangian (6) is readily extended to include 
$\eta'$--nucleon coupling.
For simplicity we work in the chiral limit.
The chiral SU(3) meson-baryon coupling Lagrangian is 
\begin{eqnarray}
{\cal L}_{\rm mB} = & &
{\rm Tr} {\overline B} (i \gamma_{\mu}
      D^{\mu} - m) B
\\ \nonumber
&+& F \
{\rm Tr} \biggl( {\overline B} \gamma_{\mu} \gamma_5 [a^{\mu}, B]_{-}
  \biggr)
+ D \ 
{\rm Tr} \biggl( {\overline B} \gamma_{\mu} \gamma_5 \{a^{\mu}, B\}_{+}
  \biggr)
\end{eqnarray}
where 
\begin{equation}
B =\
\left(\begin{array}{ccc} 
{1 \over \sqrt{2}} \Sigma^0 + {1 \over \sqrt{6}} \Lambda & \Sigma^+ & p \\
\\
\Sigma^- & -{1 \over \sqrt{2}} \Sigma^0 + {1\over \sqrt{6}} \Lambda & n \\
\\
\Xi^- & \Xi^0 & -{2 \over \sqrt{6}} \Lambda
\vphantom{\inv}  
\end{array}\right) 
\end{equation}
denotes the baryon octet and $m$ denotes the baryon mass.
In Eq.(14) $D_{\mu} = \partial_{\mu} - i v_{\mu}$ is the chiral covariant 
derivative,
$
v_{\mu} = 
-
{i \over 2} \biggl( \xi^{\dagger} \partial_{\mu} \xi +
                              \xi \partial_{\mu} \xi^{\dagger} \biggr)
$
and
$
a_{\mu} = 
-
{i \over 2} \biggl( \xi^{\dagger} \partial_{\mu} \xi -
                              \xi \partial_{\mu} \xi^{\dagger} \biggr)
$
where $\xi = U^{1 \over 2}$.
The SU(3) couplings are
$F= 0.459 \pm 0.008$ and $D= 0.798 \pm 0.008$ \cite{fec}.
The Pauli-principle forbids any flavour-singlet $J^P={1 \over 2}^+$ 
ground-state baryon degenerate with the baryon octet $B$.

We work to $O(Q^2)$ in the gluonic field and add the leading $U_A(1)$ 
invariant terms 
\begin{eqnarray}
{\cal L}_{\rm U_A(1) \ mB} = & &
{i \over 3}
K \ {\rm Tr} \biggl({\overline B} \gamma_{\mu} \gamma_5 B \biggr)
    {\rm Tr} \biggl(U^{\dagger} \partial^{\mu} U \biggr) 
\\ \nonumber
&-& {{\cal G}_{QNN} \over 2m} \partial^{\mu} Q 
  {\rm Tr} \biggl( {\overline B} \gamma_{\mu} \gamma_5 B \biggr) 
+ 
{{\cal C} \over F_0^4} Q^2 {\rm Tr} \biggl( {\overline B} B \biggr) .
\end{eqnarray}
Here, we have defined ${\cal G}_{QNN}$ and ${\cal C}$ so that 
they both have mass dimension -3. 
Some comment is warranted about the momentum scale involved in 
the coefficient multiplying the $Q^2 \ {\rm Tr} ({\bar B} B)$ term. 
There are two natural choices.
The form ${\cal C} / F_0^4$ written in Eq.(16) is motivated 
by the fact that $F_0 \sim F_{\pi}$ is relatively free of OZI 
violation.
Alternatively, motivated by the coefficient of $Q^2$ 
in Eq.(6), one 
might choose to replace ${\cal C} / F_0^4$ in Eq.(16) 
by ${\cal C}' / a F_0^2$ with ${\cal C}' = {\cal C} a / F_0^2$.
In this case
the OZI violation partially cancels in the ratio
${\cal C}' / a = 3 {\cal C}' / {\tilde m}_{\eta_0}^2$.
Note the relative size of 
the two OZI parameters ${\cal C}' \sim 33 \ {\cal C}$.
We proceed keeping the ${\cal C} / F_0^4$ 
coefficient in Eq.(16) and return to this point in Section 4.

Extra $U_A(1)$ breaking terms of the form
${\rm Tr} \biggl( \log U - \log U^{\dagger} \biggr) 
  \ Q \ {\rm Tr} \biggl( {\overline B} B \biggr)$
or
$\biggl( {\rm Tr} ( \log U - \log U^{\dagger}) \biggr)^2
{\rm Tr} \biggl( {\overline B} B \biggr)$
are excluded by the Adler-Bardeen theorem.
Such terms would correspond to extra $U_A(1)$ breaking 
in the divergence equation for $J_{\mu 5}$ 
--- in contradiction to the non-renormalisation of the 
axial anomaly in Eq.(5).
The divergence of the flavour-singlet axial-vector current 
in the effective theory 
consists of the mass terms we obtain with 
$Q={\alpha_s \over 4 \pi} G_{\mu \nu} {\tilde G}^{\mu \nu}$ 
turned off 
plus a single anomaly factor of $N_f Q$.

Since we are interested in the $pp \rightarrow pp \eta'$ 
reaction we also include possible fourth order (contact) 
terms in the baryon fields:
\begin{eqnarray}
{\cal L}_{\rm m 2B} = & &
\biggl[
\lambda_F F \
{\rm Tr} \biggl( {\overline B} \gamma_{\mu} \gamma_5 [a^{\mu}, B]_{-}
  \biggr)
+ \lambda_D D \ 
{\rm Tr} \biggl( {\overline B} \gamma_{\mu} \gamma_5 \{a^{\mu}, B\}_{+}
  \biggr)
\\ \nonumber
& & +
{i \over 3} \lambda_K 
K \ {\rm Tr} \biggl({\overline B} \gamma_{\mu} \gamma_5 B \biggr)
    {\rm Tr} \biggl(U^{\dagger} \partial^{\mu} U \biggr) 
\\ \nonumber
& & +
\lambda_Q {{\cal G}_{QNN} \over 2m} \ Q \ \partial^{\mu} 
  {\rm Tr} \biggl( {\overline B} \gamma_{\mu} \gamma_5 B \biggr) 
\biggr] 
\ {\rm Tr} \biggl( {\bar B} B \biggr)
\end{eqnarray}
where $\lambda_F$, $\lambda_D$, $\lambda_K$ and $\lambda_Q$ are 
new parameters;
${\cal L}_{\rm m2B}$ is $U_A(1)$ invariant 
for the same reason as ${\cal L}_{\rm U_A(1) \ mB}$.
Putting things together, our effective Lagrangian is
\begin{equation}
{\cal L} = 
{\cal L}_{\rm m} + {\cal L}_{\rm mB} + {\cal L}_{\rm U_A(1) \ mB}
+ {\cal L}_{\rm m 2B} .
\end{equation}

Let us consider the $Q$ dependent terms in more detail. 
The $Q$ dependent part of the effective Lagrangian (18) 
is
\begin{eqnarray}
{\cal L}_Q &=&
{1 \over 2} i \ Q \ {\rm Tr} \biggl[ \log U - \log U^{\dagger} \biggr] 
+ {1 \over a F_0^2} \ Q^2 
\\ \nonumber
&+& {{\cal G}_{QNN} \over 2 m} \ Q \
    \partial^{\mu} {\rm Tr} \biggl( {\bar B} \gamma_{\mu} \gamma_5 B \biggr) 
\ \biggl\{ 1 + \lambda_Q {\rm Tr} {\bar B} B \biggr\}
+
{{\cal C} \over F_0^4} \ Q^2 \ {\rm Tr} \biggl( {\bar B} B \biggr) ,
\end{eqnarray}
which yields the following equation of motion for $Q$:
\begin{eqnarray}
& &
{2 \over a F_0^2} \
\biggl(1 +  {{\cal C} \over F_0^2} \ a \  {\rm Tr} ( {\overline B} B )  
\biggr) \ Q 
\\ \nonumber
& & \ \ \ \ \ \ \ \ \ \ = 
- 
\biggl(
{1 \over 2} i {\rm Tr} \biggl[ \log U - \log U^{\dagger} \biggr] 
+
{{\cal G}_{QNN} \over 2m} \partial^{\mu} 
{\rm Tr} \biggl( {\overline B} \gamma_{\mu} \gamma_5 B \biggr) 
\biggl\{ 1 + \lambda_Q {\rm Tr} {\bar B} B \biggr\}
\biggl) .
\end{eqnarray}
We substitute for $Q$ in ${\cal L}_Q$ to obtain
\begin{eqnarray}
{\cal L}_Q &=& - {1 \over 12} {\tilde m}_{\eta_0}^2 \
   \biggl[ \ 6 \eta_0^2 \  
           + \ {\sqrt{6} \over m} \ {\cal G}_{QNN} \ F_0 \ 
               \partial^{\mu} \eta_0 \
           {\rm Tr} \biggl( {\bar B} \gamma_{\mu} \gamma_5 B \biggr) 
\\ \nonumber
& & \ \ \ \ \ \ \ \ 
           - \ {\cal G}_{QNN}^2 \ F_0^2 \
             \biggl( {\rm Tr} {\bar B} \gamma_5 B \biggr)^2 \
           - \ 2 \ {\cal C} \ {{\tilde m}_{\eta_0}^2 \over F_0^2 } \
           \eta_0^2 \ {\rm Tr} \biggl( {\bar B} B \biggr) 
\\ \nonumber
& & \ \ \ \ \ \ \ \ 
           + \ {\sqrt{6} \over m F_0} \ {\cal G}_{QNN} \
           \biggl\{ {1 \over 3} {\cal C} {\tilde m}_{\eta_0}^2 
                  - F_0^2 \lambda_Q \biggr\} \
           \eta_0 \ \partial^{\mu}  
           {\rm Tr} \biggl( {\bar B} \gamma_{\mu} \gamma_5 B \biggr)  \
           {\rm Tr} \biggl( {\bar B} B \biggr)
+ ... \biggr] 
\end{eqnarray}
where we have replaced $a$ by ${1 \over 3} {\tilde m}_{\eta_0}^2$.

The Lagrangian (21) has three contact terms associated with the gluonic 
potential in $Q$.
First, we observe the contact term
\begin{equation}
{\cal L}^{(1)}_{\rm contact} =
{1 \over 12} \ {\tilde m}_{\eta_0}^2 \
{\cal G}_{QNN}^2 \ F_0^2 \ \biggl( {\rm Tr} {\bar B} \gamma_5 B \biggr)^2 
\end{equation}     
in the nucleon-nucleon interaction which was discovered by Schechter et al
\cite{schechter}.
Second, we find two new contact terms associated with the axial anomaly:
\begin{equation}
{\cal L}^{(2)}_{\rm contact} =
         - {\sqrt{6} \over 12 m F_0} {\cal G}_{QNN} {\tilde m}_{\eta_0}^2 \
           \biggl\{ {1 \over 3} {\cal C} {\tilde m}_{\eta_0}^2 
                  - F_0^2 \lambda_Q \biggr\} \
           \eta_0 \ \partial^{\mu} 
           {\rm Tr} \biggl( {\bar B} \gamma_{\mu} \gamma_5 B \biggr)  \
           {\rm Tr} \biggl( {\bar B} B \biggr)
\end{equation}
and
\begin{equation}
{\cal L}^{(3)}_{\rm contact} =
  {1 \over 6 F_0^2} \ {\cal C} \ 
  {\tilde m}_{\eta_0}^4 \ \eta_0^2 \ {\rm Tr} \biggl( {\bar B} B \biggr) .
\end{equation}
The contact interaction ${\cal L}^{(2)}_{\rm contact}$ 
contributes to the low-energy $pp \rightarrow pp \eta'$ reaction;
${\cal L}^{(3)}_{\rm contact}$ 
is relevant to $\eta' N \rightarrow \eta' N$.
The three terms (22-24) describe short distance interactions.
The term ${\cal L}^{(2)}_{\rm contact}$ yields an additional 
contribution to the cross-section
for $pp \rightarrow pp \eta'$ reaction which is extra to the long 
distance contributions associated with meson exchange models 
\cite{holinde,faldt}.
The contact terms
${\cal L}^{(j)}_{\rm contact}$ 
are proportional to ${\tilde m}_{\eta_0}^2$
($j=1,2$) and
${\tilde m}_{\eta_0}^4$ ($j=3$) 
which vanish in the formal OZI limit.
Phenomenologically, the large masses of the $\eta$ and $\eta'$ 
mesons means that there is no reason, a priori, to expect the 
${\cal L}^{(j)}_{\rm contact}$ to be small.

The total contact term in the $pp \rightarrow pp \eta_0$ reaction 
is obtained by combining ${\cal L}_{\rm contact}^{(2)}$ with the 
contact terms coming from the $Q$ independent part of Eq.(17):
\begin{eqnarray}
{\cal L}^{(4)}_{\rm contact} =
           \biggl[
           &-& \sqrt{2 \over 3} 
           \biggl\{ {\lambda_D D \over F_0} + {\lambda_K K \over F_0} \biggr\}
\\ \nonumber
           &-&
           {\sqrt{6} \over 12 m F_0} {\cal G}_{QNN} {\tilde m}_{\eta_0}^2 \
           \biggl\{ {1 \over 3} {\cal C} {\tilde m}_{\eta_0}^2 
                  - F_0^2 \lambda_Q \biggr\} \
           \biggr] 
           \eta_0 \ \partial^{\mu} 
           {\rm Tr} \biggl( {\bar B} \gamma_{\mu} \gamma_5 B \biggr)  \
           {\rm Tr} \biggl( {\bar B} B \biggr)  .
\end{eqnarray}
If we substitute ${\cal C} = {\cal C}' a / F_0^2$ then the factor
$\{ {1 \over 3} {\cal C} {\tilde m}_{\eta_0}^2 - F_0^2 \lambda_Q \}$
becomes
$F_0^2 \{ {\cal C}' - \lambda_Q \}$.
We now discuss plausible values for $K$, ${\cal G}_{QNN}$, ${\cal C}$ 
and the strength of the contact term in Eq.(25).

\section{$\eta'$ nucleon coupling constants: how many ?}

The physical $\eta'$--nucleon coupling constant is read off from the 
coefficient of 
$\partial^{\mu} \eta_0 {\rm Tr} {\bar B} B$ 
in the effective Lagrangian {\it after} we have eliminated $Q$.
One finds 
\begin{equation}
g_{\eta_0 NN}
= \sqrt{2 \over 3} {m \over F_0} 
  \biggl( 2D + 2 K + {\cal G}_{QNN} F_0^2 { {\tilde m}^2_{\eta_0} \over 2 m }
\biggr) .
\end{equation}
This $g_{\eta_0 NN}$ is the $\eta_0$--nucleon coupling constant 
which will enter the extension of the coupled channels analysis 
\cite{weise} to $\eta'$ photoproduction.

The flavour-singlet Goldberger-Treiman relation for QCD was derived 
by Shore and Veneziano \cite{venez}.
In the notation of Eq.(16) one finds
\begin{equation}
m g_A^{(0)} = \sqrt{3 \over 2} F_0 \biggl( g_{\eta_0 NN} - g_{QNN} \biggr) 
\end{equation}
where
$g_{QNN} =
 \sqrt{1 \over 6} {\cal G}_{QNN} F_0 { \tilde m}^2_{\eta_0}$.
If gluonic degrees of freedom were turned off then the right hand 
side of Eq.(27) would be equal to $F_0 g_{\eta_0 NN}$.
The value of $g_A^{(0)}$ extracted from polarised deep inelastic scattering 
is \cite{bass99,windmolders}
\begin{equation}
\left. g^{(0)}_A \right|_{\rm pDIS} = 0.2 - 0.35. 
\end{equation}
The OZI prediction $g_A^{(0)} \simeq 0.6$ \cite{ej} would follow if 
polarised strange quarks and gluons were not important in the nucleon's 
internal spin structure.
{\it If} we attribute the difference between $g^{(0)}_A |_{\rm pDIS}$ 
and the OZI value 0.6 
to the gluonic correction $- \sqrt{3 \over 2} F_0 g_{QNN}$ in Eq.(27), 
{\it then} we find  $g_{QNN} \sim 2.45$ and $g_{\eta_0 NN} \sim 4.9$ 
with $F_0 \sim 0.1$GeV.
These values correspond to $K \sim -0.65$ and 
${\cal G}_{QNN} \sim +60$GeV$^{-3}$ 
if we take
${\tilde m}^2_{\eta_0} \sim 1$GeV$^2$ and substitute into Eq.(26).
The coupling constant $g_{QNN}$ is, in part,  related \cite{venez}
to the amount of spin carried by polarised gluons in a polarised 
proton.

How important is the contact interaction ${\cal L}^{(4)}_{\rm contact}$ 
in the $pp \rightarrow pp \eta'$ reaction ?

The T-matrix for $\eta'$ production in proton-proton collisions,
$p_1 (\vec{p}) + p_2 (-\vec{p}) \rightarrow p + p + \eta'$, 
at threshold in the centre of mass frame is
\begin{equation}
{\rm T^{cm}_{th}} (pp \rightarrow pp \eta') 
=
{\cal A} 
\biggl[ i ( \vec{\sigma}_1 - \vec{\sigma}_2 )
        + \vec{\sigma}_1 {\rm x} \vec{\sigma}_2 \biggr].{\vec p}
\end{equation}
where
${\cal A}$ is the (complex) threshold amplitude for $\eta'$ production.
Measurements of the total cross-section for $pp \rightarrow pp \eta'$
have been published by COSY \cite{cosy} and SATURNE \cite{saturne} at
centre of mass energies 1.5, 1.7, 2.9 and 3.7 MeV above threshold (COSY) 
and 4.1 and 8.3 MeV above threshold (SATURNE).
If taken at face value, the four COSY data points are best described by
pure three-body phase space.
If one ignores the COSY data at the two lowest energies, then the remaining
four data points are well described using the model of Bernard et al.
\cite{norbert}
treating the $pp$ final state interaction in effective range approximation.
In this model, 
which also provides a good description of $pp \rightarrow pp \pi^0$ and 
$pp \rightarrow pp \eta$ close to threshold,
one finds a best fit to the measured total cross-section data with
\begin{equation}
|{\cal A}| = 0.21 \ {\rm fm}^4
\end{equation}
with $\chi^2 = 2.4$.
The present (total cross-section only) data on
$pp \rightarrow pp \eta'$
is insufficient to distinguish between possible 
production mechanisms
involving the contact term (25) and meson exchange models.
To estimate how strong the contact term must be in order 
to make an important contribution to the measured
$pp \rightarrow pp \eta'$ cross-section,
let us consider the extreme scenario where the value of
$|{\cal A}|$ in Eq.(30) is saturated by the contact term (25).

For the values $K \simeq -0.65$ and ${\cal G}_{QNN} \simeq 60$GeV$^{-3}$
--- see below Eq.(28) ---
the contact term (25) becomes
\begin{equation}
{\cal L}_{\rm contact}^{(4)} =
\biggl(
13.1 \lambda_D - 10.6 \lambda_K - 81.6 {\cal C} + 2.4 \lambda_Q \biggr)
\ \eta_0 \  
{\rm Tr} \biggl( {\bar B} i \gamma_5 B \biggr) \ 
{\rm Tr} \biggl( {\bar B} B \biggr) . 
\end{equation}
The success of meson exchange models in describing the low-energy 
$pp \rightarrow p \Lambda K $ and $pp \rightarrow pp \eta$ 
reactions 
\cite{faldt,saturne} 
suggests that it may be reasonable to take the coefficients 
$\lambda_D$ and $\lambda_K$ of the OZI preserving processes small.
The large coefficient of ${\cal C}$ in Eq.(31) is induced by 
the OZI violation in ${\tilde m}_{\eta_0}^2$ and ${\cal G}_{QNN}$ 
(assuming that $g_{QNN}$ is indeed responsible for the small value 
of  $g_A^{(0)}|_{\rm pDIS}$ extracted from polarised deep inelastic 
scattering). 
If we saturate $|{\cal A}|$ in Eq.(30) by the OZI violating term
proportional to ${\cal C}$ then we obtain ${\cal C} \sim 1.8$GeV$^{-3}$.
Alternatively, if we use ${\tilde m}_{\eta_0}^2$ to set the scale for 
$Q^2 {\rm Tr} ({\bar B} B)$ and take 
${\cal C} / F_0^4 \mapsto {\cal C}' / a F_0^2$ 
in Eqs.(16) and (25), then ${\cal C}'$ and $\lambda_Q$ enter Eq.(31) 
with the same coefficient 2.4. 
Saturating $|{\cal A}|$ in Eq.(30) 
with the gluonic contact term proportional to 
$\{ {\cal C}' - \lambda_Q \}$ yields
$\{ {\cal C}' - \lambda_Q \} \sim 60$GeV$^{-3}$.
The OZI violating parameters
${\cal C} \sim 1.8$GeV$^{-3}$ and 
$\{ {\cal C}' - \lambda_Q \} \sim 60$GeV$^{-3}$
compare with ${\cal G}_{QNN} \sim 60$GeV$^{-3}$.

In QCD the large strange quark mass generates substantial 
$\eta$--$\eta'$ mixing.
Working in the one-mixing-angle scheme \cite{gilman} one finds 
\begin{equation}
| \eta_0 \rangle 
= \cos \theta \ | \eta' \rangle - \sin \theta \ | \eta \rangle
\end{equation}
with $\theta \simeq - 20$ degrees.
The ratio of the moduli of the contact amplitudes (25) for 
$pp \rightarrow pp \eta'$ and $pp \rightarrow pp \eta$ is 
proportional to $\cos \theta : \sin \theta = 0.94 : 0.34$.
The contact interaction (25) is likely to play a much
more prominant role in the cross-section 
for $pp \rightarrow pp \eta'$ than $pp \rightarrow pp \eta$.
In their analysis of the SATURNE data on $pp \rightarrow pp \eta'$
Hibou et al. \cite{saturne} found that a one-pion exchange model 
adjusted to fit the S-wave contribution 
to the $pp \rightarrow pp \eta$ cross-section near threshold yields 
predictions about 30\% below the measured $pp \rightarrow pp \eta'$ 
total cross-section.
The gluonic contact term (23) 
is a candidate for additional, potentially important, short range interaction.

Gluonic $U_A(1)$ degrees of freedom induce several ``$\eta'$--nucleon
coupling constants'':
$g_{\eta_0 NN}$, ${\cal G}_{QNN}$ and ${\cal C}$.
Different combinations of these coupling constants are relevant 
to different $\eta'$ 
production processes and to the flavour-singlet Goldberger-Treiman relation.
Testing the sensitivity of $\eta'$--nucleon interactions to the gluonic terms 
in the effective chiral Lagrangian for low-energy QCD 
will teach us about the role of gluons in chiral dynamics.

\vspace{0.5cm}

{\bf Acknowledgements} \\

I thank B. Borasoy, G. F\"aldt, N. Kaiser and W. Weise for helpful discussions.

\end{document}